\documentclass[fleqn,twoside]{article}
\usepackage{espcrc2,psfig}

\usepackage{graphicx}

\newcommand{\AmS}{{\protect\the\textfont2
  A\kern-.1667em\lower.5ex\hbox{M}\kern-.125emS}}

\title{Effects of large field
cutoffs in scalar and gauge models}
\author{L. Li\address[MCSD]{Department of Physics and Astronomy, \\
The University of Iowa, \\
Iowa City, Iowa 52242, USA}
       and
        Y. Meurice\addressmark[MCSD] \thanks{This 
research was supported in part by the Department of Energy
under Contract No. FG02-91ER40664.We thank the ITS of the University of Iowa for making possible
the use of its Linux clusters. 
}}
\begin{document}
\begin{abstract}
\vspace{1pc}
We discuss the notion of a large field cutoff
for LGT with compact groups. We propose and compare gauge invariant and gauge dependent (in the Landau gauge) 
criteria to sort the configurations into ``large-field'' and ``small-field'' 
configurations. We show that the correlations between volume average of field size indicators and the behavior 
of the tail of the distribution are very different in the gauge and scalar cases.
We show that the effect of discarding the large field configurations 
on the plaquette average is very different above, below and near $\beta=5.6$ for a pure $SU(3)$ LGT. 
\end{abstract}
\maketitle
A common challenge for quantum field theorists consists in 
finding accurate methods in regimes where existing expansions break down. 
In the RG language, this amounts to find acceptable interpolations for the RG flows in intermediate regions between fixed points. In a $SU(3)$ pure gauge theory 
near $\beta \simeq 6$, the validity of weak and strong coupling expansions break down and the MC method 
seems to be the only reliable method. In the following, we discuss recent attempts to 
improve weak coupling expansions.

In the case of scalar field theory, the weak coupling expansion is unable to reproduce the exponential suppression of the large field configurations coming from the factor $\exp({-\lambda\sum\phi^4_x})$ in the functional integral.
This problem can be resolved [1,2] by 
introducing a large field cutoff $|\phi|<\phi _{max}$. One is then considering a slightly different problem, however a judicious choice of $\phi _{max}$ can be used to reduce 
or eliminate the discrepancy. This optimization procedure can be approximately 
performed using the strong coupling expansion and naturally bridges the gap between 
the weak and strong coupling expansions [3]. Can such a procedure be applied to gauge models?
\begin{figure}[ht]
\vskip-25pt
\includegraphics[width=2.9in,angle=0]{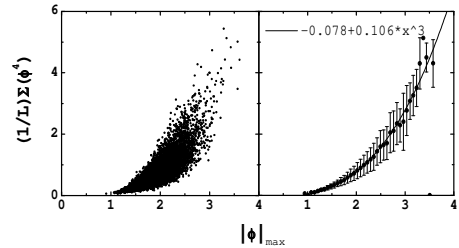}%
\vskip-45pt
\caption{Largest absolute value of the field versus average over all the sites of $\phi^4$, in $D=1$ (harmonic oscillator), for 10,000  configurations (each point corresponds to a single configuration).}
\vskip-25pt
\end{figure}

The talk of K. Wilson about the early days of lattice gauge theory 
was a very inspirational moment of this conference. He 
stressed the importance of ``butchering field theory'' in the development of the 
RG ideas and recommended that we keep doing it. 
There exist many ways to cleave 
the large field configurations. 
For scalar fields, the configurations can be ranked, for instance, according 
to the largest 
absolute value of the field  or according to the average over the sites of an even power of the field. The largest this power is, the more emphasis is put on the 
configurations with the largest field values.
We expect correlations among these quantities. 
This is illustrated for a power 4 
in the case of the harmonic oscillator in Fig. 1. 
The sample correlation is 0.82 (the maximal value being 1 for completely correlated 
quantities). In the right part of the figure, the set of configurations has been minced into 40 bins with different $\phi_{max}$. The central values can be fitted 
with a polynomial and the variance of the bins are relatively small (they would be smaller for higher correlations). Consequently, one would not expect too much change if 
one or the other method is used.

In order to understand how discarding the large field configurations changes the 
large order behavior of perturbative series, 
notice, for instance, that out of the 10,000 configurations of Fig. 1, only 56 have values of 
$|\phi|$ larger than 3. Neglecting these configurations affect the  
the order $\lambda$ correction to the ground state energy 
($\left\langle 0|\phi^4|0\right\rangle =3/4 $ without a field cut) by only 1 percent, however the same 56 
account for about 90 percent of the sixth coeffient! 

For gauge models, the closeness to the identity for a $SU(N)$ matrix $U$ 
can be measured in term of the quantity
$(1/2N)Tr \left[(1-U)(1-U)^{\dagger}\right]=1-(1/N)ReTrU$. 
Due to the compactness of the group, this quantity is bounded. For instance, it is always smaller than 2 for $SU(2)$ and 3/2 for $SU(3)$. In these two cases, the 
``largest fields'' correspond to the nontrivial elements of the center. 
For $U_L$ associated with a link $L$ in the $\mu$ direction, 
$1-(1/N)ReTrU_L\propto A_{\mu}^aA_{\mu}^a$ near the identity 
and provides an indicator of the size of the field 
which is gauge dependent. 
However, in the Landau gauge, the average of this quantity is minimized making it a prime candidate
as an indicator of field size.
On the other hand, for the product of links along a plaquette $p$, $1-(1/N)ReTrU_p$
provides a gauge invariant indicator which is proportional to the field strength near the identity.
We would like to know how these indicators are correlated.

If no gauge fixing is imposed, the distribution of $1-(1/N)ReTrU_L$ depends only on the Haar 
measure and not on $\beta$. This is illustrated in Fig. 2 for $SU(2)$ . 
However, if the Landau gauge is used, the distribution will peak much closer to zero 
as shown in Fig. 3 for $SU(3)$. The average of this quantity for a large number of 
independent configurations is 0.139 
\cite{lm}.
Note also that in the Landau gauge, there is 
a clear gap between  the maximal value taken by $1-(1/N)ReTrU_L$ (near $1.1 < 1.5$ in 
Fig. 3) and the largest possible value.

\begin{figure}[ht]
\vskip-15pt
\centerline{\psfig{figure=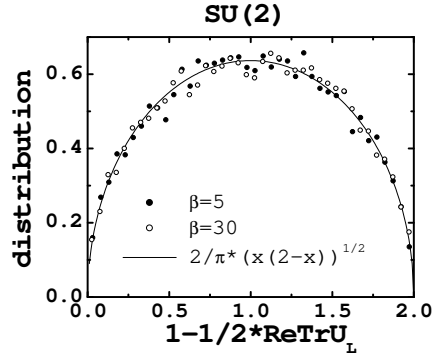,width=2.5in}}
\vskip-15pt
\caption{ Distribution of $1-\frac{1}{2}ReTrU_{L}$ in one pure $SU(2)$ configuration at various $\beta$. The 
solid line is $P(A)=\int_0^{2\pi}\frac{d   \omega}{\pi}sin^2(\frac{\omega}{2})\delta(1-\cos(\frac{\omega}{2})-A)$}
\vskip-15pt
\end{figure}
$Max_L\{1-(1/N)ReTrU_L\}$ could thus be considered as the analog of $\phi_{max}^2$ 
in the scalar case. Is this quantity correlated with field size indicators based on volume 
average as in the scalar case? 
Apparently not. The tail of the distribution of $1-(1/N)ReTrU_L$ in the Landau gauge has low population and 
may not contain relevant information about the configuration. 
It may be dependent on the algorithm used to put the configuration in the Landau gauge.
\begin{figure}[h]
\vskip-25pt
\centerline{\psfig{figure=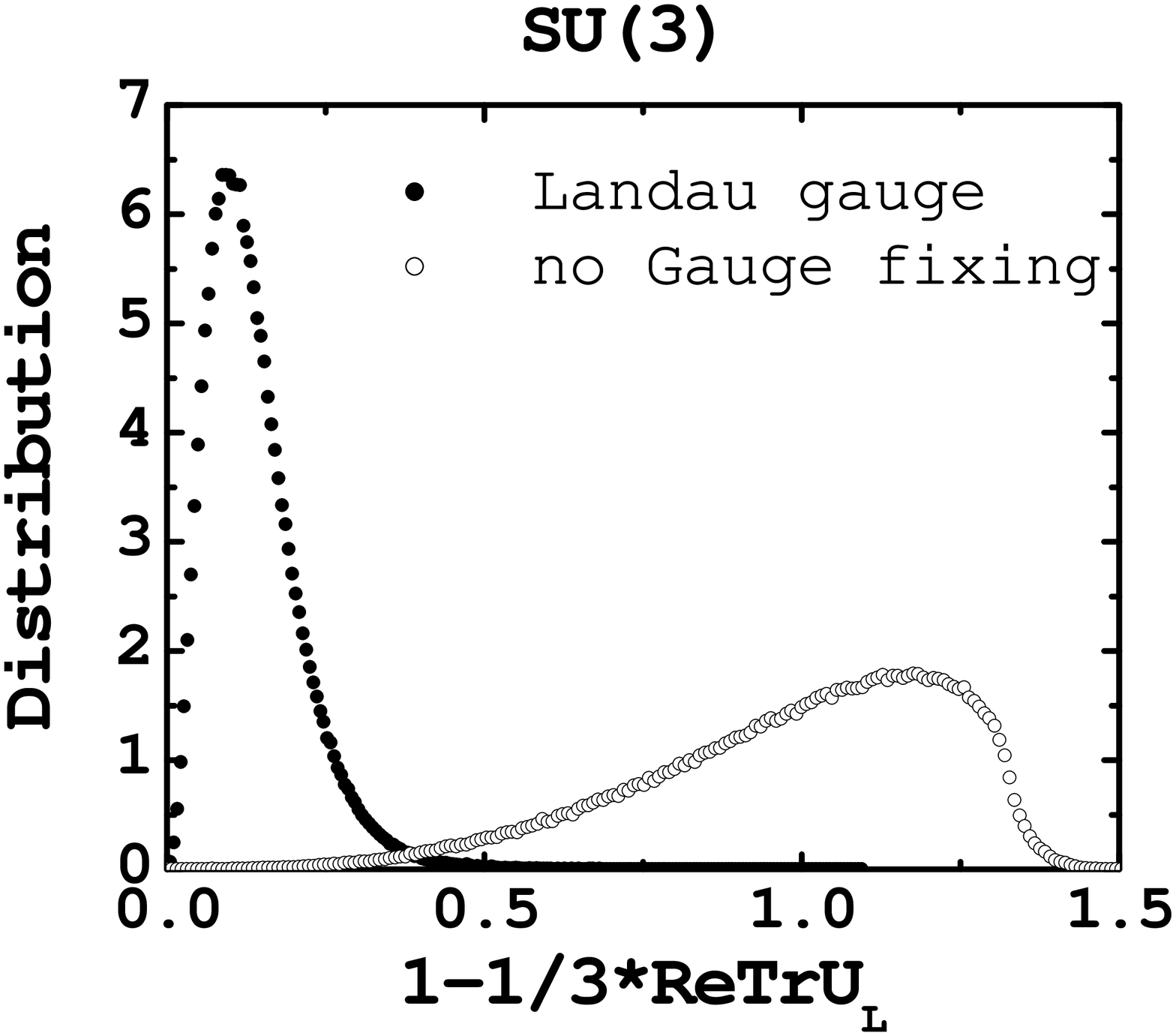,width=2.5in}}
\vskip-25pt
\caption{Distribution of $1-\frac{1}{3}ReTrU_{L}$ in one pure $SU(3)$ configuration at $\beta =6$ 
without gauge fixing (open circles) and in the Landau gauge (filled circles).}
\vskip-25pt
\end{figure}

In Fig. 4, we 
show that the maximum value 
of the distribution has practically no correlation (the sample correlation is 0.03) with 
$P=(1/6L^4)\sum_p(1-(1/N)ReTrU_p)$ 
which is a gauge invariant measure of the average size of the field of the configuration.
\begin{figure}
\centerline{\psfig{figure=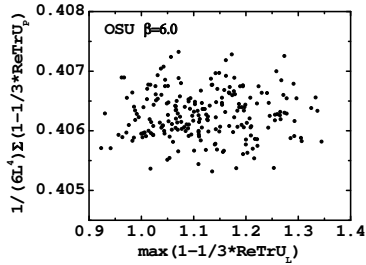,width=2in}}
\vskip-25pt
\caption{(Absence of) correlations between $Max \{1-\frac{1}{3}Re Tr U_{L}\}$ and $P$ for pure $SU(3)$ configurations at $\beta=6$ in the Landau gauge.}
\vskip-25pt
\end{figure}
On the other hand, the volume average of 
$1-(1/N)ReTrU_{L}$ in the Landau gauge is well correlated (the sample correlation is 0.46) with $P$ 
as shown in Fig. 5. After chopping the set of 221  configurations into 5 bins, the central values show a clear linear relationship. As the gauge 
invariant method is more convenient, it is a prime candidate to build modified 
perturbative series as in the scalar case. 
\begin{figure}
\vskip-5pt
\includegraphics[width=2.8in,angle=0]{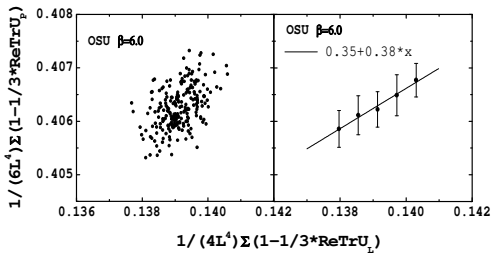}%
\vskip-30pt
\caption{Correlation between $ 1-\frac{1}{3}Re Tr U_{L}$ and $ P$ for the 
same configurations as in Fig. 4}
\vskip-25pt
\end{figure}
\begin{figure}
\centerline{\psfig{figure=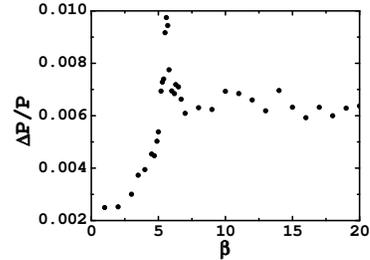,width=2in}}
\vskip-25pt
\caption{Relative change of the configuration average of $P$ when 80 percent of the large field configurations are discarded, for various values of $\beta$ in a pure $SU(3)$ LGT on a $8^4$ lattice.}
\vskip-25pt
\end{figure}

We now address the question of the dependence of an observable on a field cut relying on the 
gauge invariant criterion.
We sorted 200 independent configurations according to their value of 
$P$ and calculated the average of $P$ discarding 80 percent of the configurations with the largest values of $P$. The change in the average over configurations devided by the usual average of $P$ is 
shown in Fig. 6.
The effect of the cut is very small but of a different size below, near or above $\beta=5.6$. 
The dependence on the volume of this quantity remains to be studied. It is 
conceivable that 
$\Delta P/P$ could be taken as an order parameter.
%

The results regarding the perturbative series \cite{direnzo} for $P$ presented at the Conference 
will be discussed in a separate
preprint \cite{lili}.
Most of our calculations used FERMIQCD and MDP \cite{mdp}. The 221 configurations in the Landau gauge are from the 
public OSU configurations used in \cite{osu}.
We thank A. Gonzalez-Arroyo, M. Creutz, F. Di Renzo, M. Ogilvie, D. Sinclair and P. van Baal for valuable conversations.

\end{document}